\documentclass [6pt,a4paper]{article}
\usepackage{bbm}
\usepackage{amsfonts}
\usepackage{mathrsfs}
\usepackage {amssymb}
\usepackage {amsmath}
\usepackage{amsthm}
\usepackage{latexsym}
\def\dse#1{\vskip 0.6cm\noindent
        {\large\bf #1}
        \vskip 0.4cm}

\def\dse#1{\vskip 0.6cm\noindent
        {\large\bf #1}
        \vskip 0.4cm}
\usepackage{lastpage}
\usepackage{epsfig}

\begin{document}
\begin{center}
\textbf{\large{MacWilliams type identities on the Lee and Euclidean
weights for linear codes over $\mathbb{Z}_{\ell}$}}\footnote {
E-mail addresses: ysh$_{-}$tang@163.com(Y. Tang), sxinzhu@tom.com(S.
Zhu), kxs6@sina.com(X. Kai).}
\end{center}

\begin{center}
{Yongsheng Tang$^{1}$, Shixin Zhu$^{2}$, Xiaoshan Kai$^{2, 3}$}
\end{center}

\begin{center}
\textit{\footnotesize $^{1}$Department of  Mathematics, Hefei Normal
University, Hefei 230601, Anhui, P.R.China\\ $^{2}$School of
Mathematics, Hefei University
of Technology, Hefei 230009, Anhui, P.R.China  \\
$^{3}$National Mobile Communications Research Laboratory, Southeast
University, Nanjing 210096,  P.R.China}
\end{center}

\noindent\textbf{Abstract}~~Motivated by the works of Shiromoto [3]
and Shi et al. [4], we study the existence of MacWilliams type
identities with respect to Lee and Euclidean weight enumerators for
linear codes over $\mathbb{Z}_{\ell}.$ Necessary and sufficient
conditions for the existence of MacWilliams type identities with
respect to Lee and Euclidean weight enumerators for linear codes
over $\mathbb{Z}_{\ell}$ are given. Some examples about such
MacWilliams type identities are also presented.\\

\noindent\emph{keywords}: Linear codes,  Lee  weight enumerator,
Euclidean weight enumerator,  MacWilliams type identity

\dse{1~~Introduction} One of the most important results in coding
theory is the MacWilliams identity [2] that describes how the weight
enumerators of a linear code and its dual code relate to each other.
The identity has been found widespread applications in coding theory
and has been studied in  a lot of ways. In the 1990s, Hammons et al.
[1] found that the Lee weight of a codeword played an important role
in studying a code over $\mathbb{Z}_{4}.$ This urges that the Lee
weight enumerators of linear codes over finite rings have been
discussed by many authors. Shiromoto [3] gave the MacWilliams
identities on Lee and Euclidean weights for linear codes over
$\mathbb{Z}_{\ell}$. It is known that Shiromoto's results [3] hold
true for linear codes over $\mathbb{Z}_{4}.$ Unfortunately, these
results are not correct for more general rings. Shi et al. [4]
presented two counterexamples to Shiromoto's results [3] on the
MacWilliams type identities with respect to Lee and Euclidean weight
enumerators for linear codes over $\mathbb{Z}_{\ell}.$ However, the
authors [4] did not give the MacWilliams type identities on the Lee
and Euclidean weight enumerators for linear codes over
$\mathbb{Z}_{\ell}.$ It is natural to ask whether the MacWilliams
type identities with respect to the Lee and Euclidean weight
enumerators for linear codes over $\mathbb{Z}_{\ell}$ exist or not.
In this paper, we solve this question and give  necessary and
sufficient conditions for the existence of MacWilliams type
identities with respect to Lee and Euclidean weight enumerators for
linear codes over $\mathbb{Z}_{\ell}.$

\dse{2~~Preliminaries}

Let $\mathbb{Z}_{\ell} ( \ell  \geq2)$ denote the ring of integers
modulo $\ell$, and $\mathbb{Z}_{\ell}^{n}$ be the set of $n$-tuples
over $\mathbb{Z}_{\ell}$. A linear code $C$ of length $n$ over
$\mathbb{Z}_{\ell}$ is an additive subgroup of
$\mathbb{Z}_{\ell}^{n}$. Hence, $C$ is  a
$\mathbb{Z}_{\ell}$-submodule of $\mathbb{Z}_{\ell}^{n}$. An element
of $C$ is called a codeword of $C$. Any
$\mathbb{Z}_{\ell}$-submodule of $C$ is called a subcode of $C$.
Define the dual code $C^{\bot}$ of $C$ by
$$C^{\bot}=\left\{(x_{1},x_{2},\cdots,x_{n})\in
\mathbb{Z}_{\ell}^{n}\Big|\sum\nolimits_{i=1}^{n}x_{i}y_{i}=0,
 \  \forall (y_{1},y_{2},\cdots,y_{n})\in C\right\}.$$
Clearly, $C^{\bot}$ is also a linear code over $\mathbb{Z}_{\ell}$.
The Lee weight for the elements of $\mathbb{Z}_{\ell}$ is defined as
${\rm wt}_{{\rm L}} (a)={\rm min}\{a, \ \ell-a\}$ for all
$a\in\{0,1,\cdots,\ell-1\}$ and
$${\rm wt}_{{\rm L}}(c)=\sum\limits_{i=1}^{n}{\rm wt}_{{\rm L}}(c_{i}),$$
for  $c=(c_{1},c_{2},\cdots,c_{n})\in \mathbb{Z}_{\ell}^{n}$ (see
[5]). It is obvious that $[\ell/2]={\rm max}\{{\rm wt}_{{\rm
L}}(a)\}$ for all $a\in\{0,1,\cdots,\ell-1\},$ where $[a]$ denotes
the integer part of $a$. The Euclidean weight for the elements of
$\mathbb{Z}_{\ell}$ is defined as ${\rm wt}_{{\rm E}} (a)={\rm
wt}_{{\rm L}} (a)^{2}$ for all $a\in\{0,1,\cdots,\ell-1\}$ and
$${\rm wt}_{{\rm E}}(c)=\sum\limits_{i=1}^{n}{\rm wt}_{{\rm L}}(c_{i})^{2},$$
for  $c=(c_{1},c_{2},\cdots,c_{n})\in \mathbb{Z}_{\ell}^{n}.$ We
easily find that $[\ell/2]^{2}={\rm max}\{{\rm wt}_{{\rm E}}(a)\}$
for all $a\in\{0,1,\cdots,\ell-1\}.$ For
$c=(c_{1},c_{2},\cdots,c_{n})\in \mathbb{Z}_{\ell}^{n},$ the Hamming
weight of  $c$, denoted by ${\rm wt_{H} }(c)$, is the number of its
nonzero entries.\\

Throughout this paper, we denote by $\ell_{1}$ and $\ell_{2}$ the
following integers, respectively, $\ell_{1}=[\ell/2]$ and
$\ell_{2}=[\ell/2]^{2}.$ The Hamming weight enumerator of a linear
code $C$ of length $n$  over $\mathbb{Z}_{\ell}$ is defined as
$${\rm W}(x,y)=\sum \limits _{c\in C}x^{n-{\rm wt_{H} }(c)}y^{{\rm wt_{H} }(c)}.$$

 \noindent Clearly, ${\rm W}(x,y) =\sum \limits_{i=0}^{n}A_{i}x^{n-i}y^{i},$
where $A_{i}$ denote the number of codewords of Hamming weight $i$
in $C.$

 The Lee weight enumerator of a  linear code $C$
of length $n$ over $\mathbb{Z}_{\ell}$ is defined as

$${\rm Lee}(x,y)=\sum \limits _{c\in C}x^{\ell_{1}n-{\rm wt_{L}}(c)}y^{{\rm wt_{L}}(c)}.$$

\noindent Clearly, ${\rm Lee}(x,y) =\sum
\limits_{i=0}^{\ell_{1}n}B_{i}x^{\ell_{1}n-i}y^{i},$ where $B_{i}$
denote the number of codewords of Lee weight $i$ in $C.$

 The Euclidean  weight enumerator of a   linear code $C$ of
length $n$ over $\mathbb{Z}_{\ell}$ is defined as

$${\rm Ew}(x,y)=\sum \limits _{c\in C}x^{\ell_{2}n-{\rm wt_{E}}(c)}y^{{\rm
wt_{E}}(c)}.$$

\noindent Clearly, ${\rm Ew}(x,y) =\sum
\limits_{i=0}^{\ell_{2}n}D_{i}x^{\ell_{2}n-i}y^{i},$ where $D_{i}$
denote the
number of codewords of Euclidean weight $i$ in $C.$\\

 The following  MacWilliams identities on  Lee and
Euclidean weights
for linear codes over $\mathbb{Z}_{\ell}$ were obtained in [3]. \\

\noindent \textbf{Theorem 2.1.} Let $C$ be a  linear code of length
$n$ over $\mathbb{Z}_{\ell}$. Denote $\ell_{1}=[\ell/2]$ and
$\ell_{2}=[\ell/2]^{2}$. Then

\[
{\rm Lee}_{C^{\bot}}(x,y)= \frac{1}{|C|} {\rm
Lee}_{C}(x+(\ell^{1/\ell_{1}}-1)y,x-y);
\]

\[
{\rm Ew}_{C^{\bot}}(x,y)= \frac{1}{|C|} {\rm
Ew}_{C}(x+(\ell^{1/\ell_{2}}-1)y,x-y).
\]

For linear codes over $\mathbb{Z}_{4}$, it is known that there exist
the MacWilliams identities for Lee weight enumerators (see [1]).
That is, Theorem 2.1 can be satisfied for linear codes over
$\mathbb{Z}_{4}$. Unfortunately, it does not hold true for a general
ring $\mathbb{Z}_{l}$. This was pointed out in [4] by giving two
counterexamples. The purpose of this paper is to study the existence
of  the MacWilliams type identities with respect to the Lee and
Euclidean weight enumerators for linear codes over
$\mathbb{Z}_{\ell}.$

\dse{3~~Gray map on $\mathbb{Z}_{\ell}$}

Let $\ell$ be a fixed integer. Recall that $\ell_{1}=[ \ell/2]$. For
any element $a\in \mathbb{Z}_{\ell},$ a Gray map $\varphi$ on
$\mathbb{Z}_{\ell}$ is defined as
$$\varphi : \mathbb{Z}_{\ell}  \rightarrow  \mathbb{F}_{m}^{\ell_{1}},$$
$$ a \mapsto  (a_{1},\ldots,a_{i},a_{i+1},\ldots,
a_{\ell_{1}}),$$ where $m(>1) $ is any divisor of $\ell$ and a prime
power, and
 $\mathbb{F}_{m}$ is a finite field with $m$ elements. In detail,
 \begin{itemize}
\item if $ a=0$, then ${\rm wt}_{{\rm L}} (0)=0$ and $\varphi(a)=(0,
\ldots,0, 0,\ldots, 0);$
\item if   $0\neq a<\ell_{1}$ and ${\rm
wt}_{{\rm L}} (a)=i,$ then $\varphi(a)=(0, \ldots,0,
a_{\ell_{1}-i+1},\ldots, a_{\ell_{1}}),$ where $a_{t}\neq0$ for $
t=\ell_{1}-i+1,\ldots,\ell_{1};$
\item  if $ a=\ell_{1}$ and ${\rm wt}_{{\rm L}}
(a)=\ell_{1},$ then $\varphi(a)=(a_{1},\ldots,a_{i},a_{i+1},\ldots,
a_{\ell_{1}}),$ where $a_{t}\neq0$ for $ t=1,\ldots,\ell_{1};$
\item if $ a>\ell_{1}$ and ${\rm wt}_{{\rm L}}
(a)=i,$ then $\varphi(a)=(a_{1},\ldots,a_{i},0,\ldots, 0),$ where
$a_{t}\neq0$ for $  t=1,\ldots,i.$
 \end{itemize}
 The Gray
map $\varphi$ can be extended to $\mathbb{Z}_{\ell}^{n}$ in an
obvious way.\\

\noindent\textbf{Example 3.1.}~~Let us consider a Gray map $\varphi$
on $\mathbb{Z}_{6}.$ Since $\ell_{1}=6$, we have $\ell_{1}=3$. We
can take $m=2$ or $3$. A Gray map $\varphi$ from $\mathbb{Z}_{6}$ to
$\mathbb{F}_{m}^{3}$ can be defined as $\varphi(0)=(0,0,0)$,
$\varphi(1)=(0,0,a_{1})$, $\varphi(2)=(0,b_{2},b_{1})$,
$\varphi(3)=(c_{3},c_{2},c_{1}),$ $\varphi(4)=(d_{2},d_{1},0)$, and
$\varphi(5)=(e_{1},0,0),$ where $a_{1},b_{i},c_{j},d_{k},e_{1}\in
\mathbb{F}_{m}\setminus\{0\}.$ In particular, the Gray map $\varphi$
from $\mathbb{Z}_{6}$ to $\mathbb{F}_{2}^{3}$ can be defined as
$\varphi(0)=(0,0,0),\varphi(1)=(0,0,1),\varphi(2)=(0,1,1),\varphi(3)=(1,1,1),
\varphi(4)=(1,1,0), \varphi(5)=(1,0,0).$\\

 The following result about
the Gray map  is obvious from definition.\\
\noindent{\bf Theorem 3.2.}~~\emph{Let the notation be as before.
For any ring $\mathbb{Z}_{\ell} (\ell\geq 2),$ there exists a Gray
map $\varphi$ from $\mathbb{Z}_{\ell}^{n}$ to
$\mathbb{F}_{m}^{n\ell_{1}}$  and the Gray map $\varphi$ is a weight
preserving map from $(\mathbb{Z}_{\ell}^{n},$ Lee weight$)$ to
$(\mathbb{F}_{m}^{\ell_{1}n},$ Hamming weight$)$.}

\dse{4~~A MacWilliams type identity on Lee weight enumerator for
linear codes over $\mathbb{Z}_{\ell}$}

 For our purpose, we introduce the Krawtchouk polynomials. Let $n$ be
a fixed positive integers,  $q$  a prime power,  and $x$  an
indeterminate. The polynomials
$$K_{k}(x)=K_{k}(x,n)=\sum\limits_{j=0}^{k}(-1)^{j}(q-1)^{k-j}\binom{x}{j}\binom{n-x}{k-j},k=0,1,2,\cdots $$
are called the Krawtchouk polynomials. From definition of the
Krawtchouk polynomials, we can obtain the following two lemmas (see [2] and [6]).\\

\noindent{\bf Lemma 4.1.}~~\emph{For  non-negative integers $k$ and
$j$,$$\sum\limits_{l=0}^{n}K_{k}(l)K_{l}(j)=q^{n}\delta_{k,j},
$$ where $\delta_{k,j}=\left\{{{\begin{array}{ll}
{1,} & {if\  k = j}\\
 {0,} & {otherwise} \\\end{array} }}\right.$is the Kronecker delta.}\\

\noindent \textbf{\bf Lemma 4.2.}~~\emph{Let $C$ and $C'$ be two
codes of length $n$ over the finite field $\mathbb{F}_{q}$, and
$A_i$ and $A'_i$ be the number of codewords of weight $i$ in $C$ and
$C'$, respectively. Then
\begin{align*}
{\rm W}_{C'}(x,y)=\frac{1}{|C|}{\rm W}_{C}(x+(q-1)y,x-y).
\end{align*}\\
if and only if
\begin{align*}
A'_k=\frac{1}{|C|}\sum_{j=0}^nA_jK_k(j), k=0,1,\ldots,n.
\end{align*}
}

Let $C$ be a linear code of length $n$ over $\mathbb{Z}_{\ell},$ and
$m(>1) $ be a positive divisor of $\ell$ and a prime power. Let the
map $\varphi$ be a weight preserving map from $(
\mathbb{Z}_{\ell}^{n},$ Lee weight) to
$(\mathbb{F}_{m}^{\ell_{1}n},$ Hamming weight). Then $\varphi(C)$ is
a code of length $\ell_{1}n$ over $\mathbb{F}_m$, which is not
necessarily linear.

Let $\{A_{0},A_{1},\ldots,A_{\ell_{1}n}\}$ and ${\rm
W}_{\varphi(C)}(x,  y)$ be the Hamming weight distribution and
weight enumerator of the code $\varphi(C)$ of length $\ell_{1}n$
over $\mathbb{F}_m,$ respectively. Define their MacWilliams
transforms to be $\{A^{'}_{0},A^{'}_{1},\ldots,A^{'}_{\ell_{1}n}\}$
and ${\rm W}_{C'}(x,  y)$ of a code $C'$ of length $\ell_{1}n$ over
$\mathbb{F}_m,$  respectively. Furthermore, the MacWilliams
transforms $\{A_{0},A_{1},\ldots,A_{\ell_{1}n}\}$ and ${\rm
W}_{\varphi(C)}(x, y)$ are the Hamming weight distribution
$\{A^{'}_{0},A^{'}_{1},\ldots,A^{'}_{\ell_{1}n}\}$ and weight
enumerator ${\rm W}_{C'}(x,  y)$ of the code $C'$, respectively, and
the MacWilliams transforms of
$\{A^{'}_{0},A^{'}_{1},\ldots,A^{'}_{\ell_{1}n}\}$ and ${\rm
W}_{C'}(x,  y)$ are the weight distribution
$\{A_{0},A_{1},\ldots,A_{\ell_{1}n}\} $ and ${\rm W}_{\varphi(C)}(x,
y)$ of the code $\varphi(C),$ respectively. By Lemma 4.2, we have
\begin{align*}
A'_l=\frac{1}{|\varphi(C)|}\sum_{j=0}^{\ell_{1}n} A_jK_k(j), \ \
l=0,1,\ldots,\ell_{1}n
\end{align*}\\
and
\begin{align*}
{\rm W}_{C'}(x,y)=\frac{1}{|\varphi(C)|}{\rm
W}_{\varphi(C)}\left(x+(m-1)y,x-y\right).
\end{align*}
Moreover, for all $c \in \varphi(C),$ we have $A_{0}=1.$ By the
definition of Krawtchouk polynomials, we have  $A'_0=1.$  We know
that if $\varphi(C)$ is a linear codes of length $\ell_{1}n$ over
$\mathbb{F}_{m},$
then $C'=(\varphi(C))^{\bot}.$\\

 \noindent\textbf{\upshape Theorem 4.3.}{ \it  Let $C$ be a  linear code of length $n$ over
$\mathbb{Z}_{\ell},$ and let $m(>1) $ be a positive divisor of
$\ell$ and a prime power. Then the linear code $C$ has a MacWilliams
type identity on the Lee weight over $\mathbb{Z}_{\ell}$ with the
form

\[
{\rm Lee}_{C^{\bot}}(x,y)= \frac{1}{|C|} {\rm Lee}_{C}(x+(m-1)y,x-y)
\]
if and only if the following conditions hold true
\begin{enumerate}
\item[\emph{1)}] there exists a bijective map $\varphi$ from
$\mathbb{Z}_{\ell}^{n}$ to $\mathbb{F}_{m}^{\ell_{1}n}$ and the map
$\varphi$ is a weight preserving map from $( \mathbb{Z}_{\ell}^{n},$
Lee weight) to $(\mathbb{F}_{m}^{\ell_{1}n},$ Hamming weight);
\item[\emph{2)}] there exists  a code $C'$ of length
$\ell_{1}n$ over $\mathbb{F}_m$  and the  MacWilliams transform
  ${\rm W}_{C'}(x,y)$ of ${\rm W}_{\varphi(C)}(x,y)$ satisfying ${\rm
W}_{\varphi(C^{\bot})}(x,  y)={\rm W}_{C'}(x,  y)$.
\end{enumerate}}

\noindent \textbf{Proof.} First, suppose that  a  linear code $C$ of
length $n$ has a MacWilliams type identity on the Lee weight over
$\mathbb{Z}_{\ell}$ with the form

\[
{\rm Lee}_{C^{\bot}}(x,y)= \frac{1}{|C|} {\rm
Lee}_{C}(x+(m-1)y,x-y).
\]
 By Theorem 3.2, we
attain that $\varphi(C)$ is  a code of length $\ell_{1}n$ over
$\mathbb{F}_{m}$ and
\[
{\rm Lee}_{C}(x,y)= {\rm W}_{\varphi(C)}(x,y).
\]
For the code $\varphi(C),$ there exists  a code $C'$ of length
$\ell_{1}n$ over $\mathbb{F}_m$  and the  MacWilliams transform
  ${\rm W}_{C'}(x,y)$ of ${\rm W}_{\varphi(C)}(x,y)$ satisfying
\[
{\rm W}_{C'}(x,y)= \frac{1}{|\varphi(C)|} {\rm
W}_{\varphi(C)}(x+(m-1)y,x-y).
\]
Furthermore
\[
{\rm Lee}_{C^{\bot}}(x,y)= {\rm W}_{\varphi(C^{\bot})}(x,y)
=\frac{1}{|C|} {\rm W}_{\varphi(C)}(x+(m-1)y,x-y).
\]
It follows that
\begin{eqnarray}
|C| {\rm W}_{\varphi(C^{\bot})}(x,y) =|\varphi(C)|{\rm W}_{C'}(x,y).
\end{eqnarray} Note that $A_{0}=1$ and $A'_0=1$. By comparing
the coefficient of $x^{\ell_{1}n}$ in the R.H.S. of Equation (1)
with the L.H.S. of Equation (1), we obtain
$$|C|=|\varphi(C)|.$$ Therefore
$${\rm W}_{\varphi(C^{\bot})}(x,  y)={\rm W}_{C'}(x,  y).$$
This shows that the  conditions 1) and 2) hold true.

 On the other hand, if
there exists a bijective map $\varphi$ from $\mathbb{Z}_{\ell}^{n}$
to $\mathbb{F}_{m}^{\ell_{1}n}$ and the map $\varphi$ is a weight
preserving map from $( \mathbb{Z}_{\ell}^{n},$ Lee weight) to
$(\mathbb{F}_{m}^{\ell_{1}n},$ Hamming weight), then
\[
{\rm Lee}_{C}(x,y)= {\rm W}_{\varphi(C)}(x,y)
\] and $$|C|=|\varphi(C)|.$$
Furthermore, for the code $\varphi(C),$ there exists  a code $C'$ of
length $\ell_{1}n$ over $\mathbb{F}_m$  and the  MacWilliams
transform  ${\rm W}_{C'}(x,y)$ of ${\rm W}_{\varphi(C)}(x,y)$
satisfying
\[
{\rm W}_{C'}(x,y)= \frac{1}{|\varphi(C)|} {\rm
W}_{\varphi(C)}(x+(m-1)y,x-y).
\]
Since ${\rm W}_{\varphi(C^{\bot})}(x,  y)={\rm W}_{C'}(x, y)$, then
\[
{\rm W}_{\varphi(C^{\bot})}(x,y)= \frac{1}{|\varphi(C)|} {\rm
W}_{\varphi(C)}(x+(m-1)y,x-y).
\]
Therefore
\[
{\rm Lee}_{C^{\bot}}(x,y)= \frac{1}{|C|} {\rm
Lee}_{C}(x+(m-1)y,x-y). \] \qed\\

\noindent\textbf{Remark }~~If the code $\varphi(C)$ is  a linear
code of length $\ell_{1}n$ over $\mathbb{F}_{m},$    then
$C'=(\varphi(C))^{\bot}$ in the condition 2) of Theorem 4.3. \\

 From Theorem 4.3,  we  easily get a necessary and sufficient condition for the existence of the
MacWilliams type identities on the Lee and Euclidean weight
enumerators for linear codes over $\mathbb{Z}_{\ell}.$\\

\noindent\textbf{\upshape Corollary 4.4.}~~{\it Let $C$ be a linear
code of length $n$ over $\mathbb{Z}_{\ell},$  and let $m(>1) $ be a
positive divisor of $\ell$ and a prime power. Then the linear code
$C$ has a MacWilliams type identity on the Lee weight over
$\mathbb{Z}_{\ell}$ with the form

\[
{\rm Lee}_{C^{\bot}}(x,y)= \frac{1}{|C|} {\rm Lee}_{C}(x+(m-1)y,x-y)
\]
if and only if $\ell=m^{\ell_{1}}$ and there exists  a code $C'$ of
length $\ell_{1}n$ over $\mathbb{F}_m$  and the  MacWilliams
transform
  ${\rm W}_{C'}(x,y)$ of ${\rm W}_{\varphi(C)}(x,y)$ satisfying ${\rm
W}_{\varphi(C^{\bot})}(x,  y)={\rm W}_{C'}(x,  y)$. }\\

In fact, Corollary  4.4 gives a criterion for judging the existence
of the MacWilliams type identities on the Lee  weight
enumerator for linear codes over $\mathbb{Z}_{\ell}.$ Using this criterion we obtain the following result.\\

 \noindent \textbf{\upshape Corollary 4.5.}~~{ \it  Let $C$ be a linear
code of length $n$ over $\mathbb{Z}_{\ell}(\ell\geq 5),$  and let
$m(>1) $ be a positive divisor of $\ell$ and a prime power. Then
there is no  MacWilliams type identity on the Lee weight for the
linear code $C$ over $\mathbb{Z}_{\ell}$ with the form

\[
{\rm Lee}_{C^{\bot}}(x,y)= \frac{1}{|C|} {\rm
Lee}_{C}(x+(m-1)y,x-y).
\]}\\
\noindent \textbf{Proof.} By Corollary 4.4, we have $m=\ell^{1/
\ell_{1}}.$ Now, we prove the result by considering three cases.

\begin{enumerate}
\item[(i)] $\ell=5.$  Then $\ell_{1}=2.$ Since
$m=\sqrt{5}$ is not a  positive integer, then there is no bijective
map $\varphi$ from $\mathbb{Z}_{5}^{n}$ to $\mathbb{F}_{m}^{2n}.$
\item[(ii)] $\ell \geq6$ is even. Denote $\ell=2\kappa.$
Then $\kappa \geq 3$ and $\ell_{1}=\kappa.$   Let
$\sqrt[\kappa]{2\kappa}=t+1.$ Then we have
$2\kappa=(t+1)^{\kappa}>\kappa t+\frac{\kappa(\kappa-1)}{2}t^{2}.$
It follows that $2>t+t^{2},$ which means $t<1.$  Therefore
$1<m=\sqrt[\kappa]{2\kappa}<2.$ This contradicts the fact that
$m(>1)$ is a positive divisor of $\ell.$
\item[(iii)] $\ell > 6$ is odd. Denote $\ell=2\kappa+1.$
Then $\kappa \geq 3$ and $\ell_{1}=\kappa.$   Let
$\sqrt[\kappa]{2\kappa+1}=t+1.$ Similar to Case 2, we can get a
contradiction.\qed
\end{enumerate}

Let us use the above results to consider a  linear code $C$ of
length $n(\geq 1)$ over $\mathbb{Z}_{4}$ on the Lee weight. First,
there exists a bijective map $\varphi$ from $\mathbb{Z}_{4}$ to
$\mathbb{F}_{2}^{2}.$ In fact, $\varphi(0)=(0,0)$,
$\varphi(1)=(0,1)$, $\varphi(2)=(1,1)$, and $\varphi(3)=(1,0),$ and
$\varphi(C)$ is nonlinear(see [1] and [6]). The  map $\varphi$ can
be extended to $\mathbb{Z}_{4}^{n}$ in an obvious way and the
extended $\varphi$ is a bijection from $\mathbb{Z}_{4}^{n}$ to
$\mathbb{F}_{2}^{2n}.$ Second, there exists  a code $C'$ of length
$2n$ over $\mathbb{F}_2$  and the  MacWilliams transform
  ${\rm W}_{C'}(x,y)$ of ${\rm W}_{\varphi(C)}(x,y).$
By Lemma 4.2, we have ${\rm W}_{C'}(x, y)=\frac{1}{|\varphi(C)|}
{\rm W}_{\varphi(C)}(x+y, x-y).$ Then, we get
$A'_l=\frac{1}{|\varphi(C)|}\sum_{j=0}^{2n}A_jK_k(j).$ By Lemma 4.1,
we have
\begin{eqnarray*}
\frac{1}{|C'|}\sum_{l=0}^{2n}A'_lK_k(l)=&&\frac{1}{|C'|}\frac{1}{|\varphi(C)|}\sum_{l=0}^{2n} \big(\sum_{j=0}^{2n}A_jK_l(j)\big)K_k(l)\\
=&&\frac{1}{|\varphi(C)||C'|}\sum_{j=0}^{2n} A_j \sum_{l=0}^n K_l(j)K_k(l) \\
=&&\frac{1}{|\varphi(C)||C'|}\sum_{j=0}^{2n} A_j 2^{2n}\delta_{j,k}. \\
\end{eqnarray*}
Therefore, $\frac{1}{|C'|}\sum_{l=0}^{2n}A'_lK_k(l)=A_j$ if and only
if $\frac{1}{|\varphi(C)||C'|}\sum_{j=0}^{2n} A_j
2^{2n}\delta_{j,k}=A_j,$ that is, $|\varphi(C)||C'|=2^{2n}.$  On the
other hand, for the code $C',$ there exists a linear code
$\mathcal{C}$ over $\mathbb{Z}_{4}$ such that
$C'=\varphi(\mathcal{C}).$  Then,
$|\varphi(C)||C'|=|\varphi(C)||\varphi(\mathcal{C})|=2^{2n}.$ Since
$\varphi$ is a bijection, it follows that
$|\varphi(C)||\varphi(\mathcal{C})|=|C||\mathcal{C}|=2^{2n}.$
Therefore, $\varphi(\mathcal{C})=(\varphi(C))^{\bot}$ or
$\mathcal{C}=C^{\bot}$ (see [7]). Since  $\varphi(C)$ is nonlinear,
 it only follows that  $\mathcal{C}=C^{\bot}.$  Then  the MacWilliams
transform ${\rm W}_{C'}(x,y)$ of ${\rm W}_{\varphi(C)}(x,y)$
satisfying ${\rm W}_{\varphi(C^{\bot})}(x,  y)={\rm W}_{C'}(x,  y)$.
Hence, ${\rm W}_{\varphi(C^{\bot})}(x, y)=\frac{1}{|\varphi(C)|}
{\rm W}_{\varphi(C)}(x+y, x-y).$ Finally, the linear code $C$ of
length $n$ over $\mathbb{Z}_{4}$ has MacWilliams type identity on
the Lee weight with the form

\[
{\rm Lee}_{C^{\bot}}(x,y)= \frac{1}{|C|} {\rm Lee}_{C}(x+y,x-y).
\]

The following two examples demonstrate the non-existence of
MacWilliams type identities on the Lee weight for linear codes over
$\mathbb{Z}_{6}$ and $\mathbb{Z}_{8},$ respectively.\\

\noindent \textbf{ \bf Example 4.6.}~~Consider any  linear code $C$
of length $n(\geq 1)$ over $\mathbb{Z}_{6}$ equipped with the Lee
weight. Since there does not exist  a bijective map $\varphi$ from
$\mathbb{Z}_{6}^{n}$ to $\mathbb{F}_{m}^{3n}$ ( $m=2$ or $3$), the
linear code $C$ of length $n$ over $\mathbb{Z}_{6}$ does not have a
MacWilliams type identity on the Lee weight with the form

\[
{\rm Lee}_{C^{\bot}}(x,y)= \frac{1}{|C|} {\rm
Lee}_{C}(x+(m-1)y,x-y).
\]

\noindent \textbf{ \bf Example 4.7.}~~Consider any  linear code $C$
of length $n(\geq 1)$ over $\mathbb{Z}_{8}$ equipped with the Lee
weight. Since there does not exist  a bijective map $\varphi$ from
$\mathbb{Z}_{8}^{n}$ to $\mathbb{F}_{m}^{4n}$ ( $m=2,4$ or $8$),
then the  linear code $C$  of length $n$ over $\mathbb{Z}_{8}$ does
not have a  MacWilliams type identity on the Lee weight with the
form

\[
{\rm Lee}_{C^{\bot}}(x,y)= \frac{1}{|C|} {\rm
Lee}_{C}(x+(m-1)y,x-y).
\]

\dse{5~~A MacWilliams type identity on Euclidean weight enumerator
for linear codes over $\mathbb{Z}_{\ell}$}

In this section, we will use the similar methods in Section 4 to
study the MacWilliams type identity on the Euclidean weight
enumerator for linear codes over $\mathbb{Z}_{\ell}.$ For every
element $a\in \mathbb{Z}_{\ell},$ a map $\Phi$ on
$\mathbb{Z}_{\ell}$ is defined as
$$\Phi : \mathbb{Z}_{\ell}  \rightarrow  \mathbb{F}_{q}^{\ell_{2}},$$
$$ a \mapsto  (a_{1},\ldots,a_{i},a_{i+1},\ldots,
a_{\ell_{2}}),$$ where $q(>1)$ is a positive divisor of $\ell$ and a
prime power, and $\mathbb{F}_{q}$ is a finite field with $q$
elements. Similar to Theorem 4.3, we can obtain the following
results.

\noindent\textbf{\upshape Theorem 5.1.}~~{ \it  Let $C$ be a linear
code of length $n$ over $\mathbb{Z}_{\ell}.$   Let $q(>1) $ be a
positive divisor of $\ell,$ and a prime power. Then the  linear code
$C$ has a MacWilliams type identity on the Euclidean weight over
$\mathbb{Z}_{\ell}$ with the form

\[
{\rm Ew}_{C^{\bot}}(x,y)= \frac{1}{|C|} {\rm Ew}_{C}(x+(q-1)y,x-y)
\]
if and only if the following conditions hold
true:\\
1) there exists a bijective map $\Phi$ from $\mathbb{Z}_{\ell}^{n}$
to $\mathbb{F}_{q}^{\ell_{2}n}$ and the map $\Phi$ is a weight
preserving map from $( \mathbb{Z}_{\ell}^{n},$ Euclidean weight) to
$(\mathbb{F}_{q}^{\ell_{2}n},$
Hamming weight)  ;\\
2)there exists  a code $C''$ of length $\ell_{2}n$ over
$\mathbb{F}_q$  and the  MacWilliams transform
  ${\rm W}_{C''}(x,y)$ of ${\rm W}_{\Phi(C)}(x,y)$ satisfying ${\rm
W}_{\Phi(C^{\bot})}(x,  y)={\rm W}_{C''}(x,  y)$. }\\

\noindent\textbf{\upshape Corollary 5.2.}~~{ \it  Let $C$ be a
linear code of length $n$ over $\mathbb{Z}_{\ell}.$   Let $q(>1) $
be a positive divisor of $\ell$ and a prime power. Then the  linear
code $C$ has a MacWilliams type identity on the Euclidean weight
over $\mathbb{Z}_{\ell}$ with the form

\[
{\rm Ew}_{C^{\bot}}(x,y)= \frac{1}{|C|} {\rm Ew}_{C}(x+(q-1)y,x-y)
\]
if and only if $\ell=q^{\ell_{2}}$ and there exists  a code $C''$ of
length $\ell_{2}n$ over $\mathbb{F}_q$  and the  MacWilliams
transform
  ${\rm W}_{C''}(x,y)$ of ${\rm W}_{\Phi(C)}(x,y)$ satisfying ${\rm
W}_{\Phi(C^{\bot})}(x,  y)={\rm W}_{C''}(x,  y)$.}\\

By using the above corollary, we can obtain the nonexistence of a
MacWilliams type identity on the Euclidean weight for  linear codes
over $\mathbb{Z}_{\ell}$  once the integer $\ell$ is more
than $3$.\\

\noindent \textbf{\upshape Corollary 5.3.}~~{ \it  Let $C$ be a
linear code of length $n$ over $\mathbb{Z}_{\ell}(\ell\geq 4).$ Let
$q(>1) $ be a positive divisor of $\ell$ and a prime power.  There
is no  MacWilliams type identity on the Euclidean weight for the
linear code $C$ over $\mathbb{Z}_{\ell}$ with the form

\[
{\rm Ew}_{C^{\bot}}(x,y)= \frac{1}{|C|} {\rm Ew}_{C}(x+(q-1)y,x-y).
\]}\\
\noindent \textbf{Proof.} By Corollary 5.2, we have $q=\ell^{1/
\ell_{2}}.$ Now, we divide into two cases to prove the reult.

\begin{enumerate}
\item[(i)] $\ell \geq4$ is even. Denote $\ell=2\rho.$
Then $\rho \geq 2$ and $\ell_{2}=\rho^{2}.$   Let
$\sqrt[\rho^{2}]{2\rho}=\lambda+1.$ Then
$2\rho=(\lambda+1)^{\rho^{2}}>\rho^{2}\lambda+\frac{\rho^{2}(\rho^{2}-1)}{2}\lambda^{2}.$
It follows that $2>\lambda+\lambda^{2},$ which gives $\lambda<1.$
Therefore $1<q=\sqrt[\rho^{2}]{2\rho}<2,$ which contradicts the fact
that $q(>1) $ is a positive divisor of $\ell.$
\item[(ii)] $\ell > 4$ is odd. Denote $\ell=2\rho+1.$
Then $\rho \geq 2$ and $\ell_{2}=\rho^{2}.$   Let
$\sqrt[\rho^{2}]{2\rho+1}=\lambda+1.$ Then we have
$2\rho+1=(\lambda+1)^{\rho^{2}}>1+\rho^{2}\lambda+\frac{\rho^{2}(\rho^{2}-1)}{2}\lambda^{2}.$
It follows that $2>\lambda+\lambda^{2}$, which means $\lambda<1.$
Therefore $ 1<q=\sqrt[\rho^{2}]{2\rho+1}<2,$ which contradicts the
fact that $q(>1) $ is a positive divisor of $\ell$. \qed
\end{enumerate}

Corollaries 4.5 and 5.2 give necessary and sufficient conditions for
the existence of Mac-Williams type identities on the Lee and
Euclidean weight for linear codes over $\mathbb{Z}_{\ell}$,
respectively. From them we can see that Theorem 2.1 does not always
hold true for all positive integer $\ell$. The existence of
MacWilliams type identities on the Lee and Euclidean weight
enumerators for linear codes over $\mathbb{Z}_{\ell}$ depends on the
value of $\ell$ and Gray map.

\dse{~~Acknowledgements} This research is supported by National
Natural Science Funds of China (Nos. 61370089 and 61572168), Natural
Science Foundation of Anhui Province (No. 1408085QF116), National
Mobil Communications Research Laboratory, Southeast University( No.
2014D04), Colleges Outstanding Young Talents Program in 2014, Anhui
Province ( No. [2014]181), Anhui Province Natural Science Research
(No. KJ2015A308) and Hefei Normal University Research Project (No.
2015JG09). The authors would like to thank the anonymous referees
who gave many helpful suggestions and comments to greatly improve
the presentation of the paper.

\end{document}